\documentstyle[amssymb,preprint,aps]{revtex}

\begin{document}
\title{On X-Ray Channeling in $\mu$- and $n$-capillaries}
\author{S. Bellucci $^{1}$\thanks{%
Corresponding author: E-mail: bellucci@lnf.infn.it; Phone: +39 (06) 9403
2888; Fax: +39 (06) 9403 2427}, S.B. Dabagov $^{1,2}$}
\address{1) - INFN - Laboratori Nazionali di Frascati, P.O. Box 13, I-00044 Frascati,%
\\
Italy \\
2) - RAS - P.N. Lebedev Physical Institute, 119991 Moscow, Russia}
\maketitle

\begin{abstract}
In this work X-ray propagation in micro- and nano-size capillaries has been
considered in the frame of a simple unified wave theory. It is shown that
the diminishing of the channel sizes completely changes the mode of beam
transportation, namely, we obtain the transformation of surface channeling
in microcapillaries to bulk channeling in nanocapillaries (nanotubes).
\end{abstract}



\section{{\bf INTRODUCTION}\protect\bigskip\noindent}

\noindent Starting from the mid 1980s the interest for the problem of X-ray
passage through capillary systems increased, because of the development of a
new technology for the manufacturing of X-ray optics of grazing incidence
\cite{PhRep}. Nowadays capillary optics represents a well-established X-ray
and neutron optical instrument that allows experiments to be performed quite
efficiently on a much smaller scale than conventional X-ray devices. These
optical elements consist of hollow tapered tubes that condense neutral
particles by multiple reflections from the inner channel surface \cite
{Engstrom,Stern}. \noindent Moreover, capillary optics relies also on the
ability of a tapered and/or bent capillary channel to act as an X-ray
waveguide \cite{spiseg-apl74}, in other words, an optical element may be
considered as a whispering-gallery X-ray device.\noindent\

\noindent As well known, a whispering gallery device is a multiple
reflection one with a large number of bounces \cite{vino1}. The total
subtended angular aperture of the device is determined by the number of
bounces $N$ and the glancing angle $\theta$, and equals $2N\theta$. While
the reflectivity of single bounce light from a substance with the complex
dielectric function $\varepsilon=1-\delta+i\beta$ ($\delta$ and $\beta$ are
the parameters of polarizability and attenuation, respectively) is defined by

\begin{equation}
R_{1}%
{s \atopwithdelims[] p}%
\simeq1-2%
\mathop{\rm Re}%
\left(
{1 \atopwithdelims[] \varepsilon}%
\left( \varepsilon-1\right) ^{-1/2}\right) \ ,
\end{equation}

\noindent where $s,p$ are the indexes of\ radiation polarization, the
integral reflection may be estimated by taking the $\theta\rightarrow0$
limit -

\begin{equation}
R%
{s \atopwithdelims[] p}%
\simeq\exp\left( -R_{1}%
{s \atopwithdelims[] p}%
N\theta\right)
\end{equation}

\noindent Evaluations made by these relations have shown that the whispering
galleries offer high efficiency with narrow band pass\cite{smho-nima94}.

\noindent Application of wave theory to the radiation passage through
capillary structures opens up new prospects for the study\cite
{da-rep92,daku-spie95,alda-nim98}. As will be shown below, capillary optical
systems act in such a way that radiation propagating in channels consists of
two portions: one is scattered by the laws of ray optics, the other one is
captured in bound modes by a surface potential. Moreover, when the channel
size values approach the transverse wavelength of radiation, the bulk
channeling of photons occurs similarly to the channeling of charged
particles in crystals\cite{Lindhard}. At present, the technology of
capillary system manufacturing allows structures of the deep submicron level
to be produced \cite{kum-selpap00}. For example, as examined samples one may
consider the carbon nanostructures (nanocapillary systems) \cite{buda-nc01},
in the fabricating of which a big progress has been achieved in recent years
\cite{3}.

A graphitic nanotube \cite{iij-nat91,ajiij-nat93,iijichi-nat93} may be
considered as a small-size (nano-size) capillary. The wall surface of it is
formed by carbon atoms, the distance among which is estimated to be $1\div 2$
\AA . The typical diameter of a single nanotube is about tens of nanometers
(the channel diameter / wall thickness ratio may reach two orders of
magnitude), and the length of such a structure may be of submillimeter
order. All these features of the nanotube structures result in the
utilization of wave propagation theory instead of the ray approximation one,
and consequently, nanotubes may be considered as X-ray waveguides. This
allows the well known channeling theory to be applied for describing X-ray
beam propagation inside these structures \cite{zhevago,dedkov}.

Nanotube research has been developed rapidly over the last decade following
the bulk production of C60 and structural identification of carbon
nanotubes. The discovery of a third carbon allotrope Buckminsterfullerene
(C60) in the mid 1980s initiated and led to the development of elongated
cage-like structures (known as nanotubes) in 1990s. Since 1991 \cite{1},
there has been a lot of study on carbon nanotubes to understand their
formation and properties. In 1992, reports appeared on the bulk synthesis of
nanotubes formed as an inner core cathode deposit generated by arching
graphite electrodes in an inert atmosphere. The products exhibit different
degrees of crystallinity and various morphologies (e.g. straight, curled,
hemitoroidal, branched, spiral, helix shaped etc.).

Indeed, carbon nanotubes stick out in the field of nanostructures, owing to
their exceptional mechanical, capillarity, electronic transport and
superconducting properties \cite{2}. They are cylindrical molecules with a
diameter of order 1 nm and a length of many microns. They are made of carbon
atoms and can be thought of as a graphene sheet rolled around a cylinder
\cite{4}.

It is well known that nanotubes can be manufactured of different diameters -
from a fraction of nm up to microns, of different length - from tens of
microns up to millimeters, of different materials - usually carbon but also
others \cite{5}. There are various applications of nanotubes. Recent
examples include use of nanotube as (I) gas storage components for Ar, N2
and H2 (II) STM probes and field emission sources (III) high power
electrochemical capacitors (IV) chemical sensors (V) electronic nanoswitches
(VI) magnetic storage devices etc.

Different carbon nanotubes have been manufactured and investigated by X-ray
absorption near-edge structure (XANES) at the carbon K edge. The surface
structure and bonding properties have been studied by characteristic
pre-edge features while structural features of these nanostructured systems
have been identified looking at the variation in the multiple-scattering
region of the XANES spectra \cite{6}.

Simulations of particle beam channeling in carbon nanotubes have been
recently performed, in order to evaluate the possibilities for experimental
observation of channeling effects in both straight and bent nanotubes,
considering different charged particle species, such as protons of 1.3 and
70 GeV, and positrons of 0.5 GeV \cite{7}. There the capabilities of a
nanotube channeling technique for charged particle beam steering have been
discussed, based on earlier Monte Carlo simulations \cite{8}.

In this work a unified description for X-ray propagation (note that the
same theory is valid for neutrons) through capillaries of various diameters
is presented.

\section{QUANTUM-WAVE DUALISM}

\noindent\noindent As it has been shown recently, the propagation of X-ray
photons through capillary systems exhibits a rather complex character \cite
{da...jsr95}. Not all features shown in the experiments can be explained
within the geometrical (ray) optics approximation \cite
{arar-phscr98,mono,kukh.etal-nimb00}. On the contrary, the application of
wave optics methods allows us to describe in details the processes of
radiation spreading into capillaries.

\noindent The propagation of X radiation through capillary systems is mainly
defined by its interaction with the inner channel walls. In the ideal case,
when the boundary between hollow capillary and walls represents a smooth
edge, the beam is split in two components: the mirror-reflected and
refracted ones. The latter appears sharply suppressed in the case of total
external reflection. The characteristics of scattering inside capillary
structures can be evaluated from the solution of a wave propagation
equation. In the first order approximation, without taking into account the
roughness correction $\Delta\varepsilon\left( {\bf r}\right) =0$ ($%
\Delta\varepsilon$ is the perturbation in dielectric permittivity induced by
the presence of roughness), the wave equation in the transverse plane to the
propagation direction reads

\begin{equation}
\left( -\nabla^{2}+k^{2}\delta\left( {\bf r}_{\perp}\right) -k_{\perp
}^{2}\right) E\left( {\bf r}_{\perp}\right) =0\ ,
\end{equation}

\noindent where $E$ is a function of the radiation field, and ${\bf k\equiv (}%
k_{\parallel },k_{\perp }{\bf )}$ is a wave vector.

\noindent Due to the fact that the transverse wave vector may be presented
as $k_{\perp}\approx k\theta$ under the grazing wave incidence ($\theta\ll1$%
), an ''effective interaction potential'' is estimated by the expression

\begin{equation}
V_{eff}\left( {\bf r}_{\perp}\right) =k^{2}\left( \delta\left( {\bf r}%
_{\perp}\right) -\theta^{2}\right) =\left\{
\begin{array}{rcl}
-k^{2}\theta^{2} & , & r_{\perp}<r_{1} \\
k^{2}\left( \delta_{0}-\theta^{2}\right) & , & r_{\perp}\geq r_{1}\,\,,
\end{array}
\right.  \label{e-pot-flat}
\end{equation}
where $r_{1}$ corresponds to the reflecting wall position. From the latter
the phenomenon of total external reflection at $V_{eff}=0$ follows, when $%
\theta\equiv\theta_{c}\simeq\sqrt{\delta_{0}}$ - the Fresnel's angle.

\noindent When we introduce a curvature in the reflecting surface, the
effective potential obtains an additional contribution. This term, that
corresponds to the additional ''potential energy'', can be seen physically in
the following way. Due to the reflecting surface curvature a photon receives
an angular momentum $kr_{curv}\varphi $ of the ''centripetal force'', where $%
r_{curv}$ is a curvature radius of the photon trajectory. The latter is
supplied by the ''centrifugal potential energy'' $-k^{2}r_{\perp }/\left(
2r_{curv}\right) $

\begin{equation}
V_{eff}\left( {\bf r}_{\perp}\right) =k^{2}\left( \delta\left( {\bf r}%
_{\perp}\right) -\theta^{2}-\frac{r_{\perp}}{2r_{curv}}\right) .
\label{e-pot-curv}
\end{equation}

\noindent The situation is explained in the scheme in Fig. 1. Because of the
variation in the spatial system parameters, the interaction potential has
been changed from the step potential with the potential barrier of $%
k^{2}\delta _{0}$ to the well potential, with the depth and width defined by
the channel characteristics.

\noindent In the following we briefly discuss a solution of the wave
equation in the case of an ideal reflecting surface (i.e. without
roughness), when the reflected beam is basically determined by the
coherently scattered part of radiation (for details see Refs. \cite
{features,coh-incoh}). The evaluation of the wave equation with the boundary
conditions of a capillary channel shows that X-radiation may be distributed
over the bound state modes defined\ by the capillary channel potential (see
below). It is important to underline here that the channel potential acts as
an effective specular reflecting barrier, and then, the effective
transmission of X-radiation by the hollow capillary tubes is observed. While
the main portion of radiation undergoes the incoherent diffuse scattering,
the remaining contribution (usually small) is due to coherent scattering
that represents a special phenomenon, extremely interesting to observe and
clarify \cite{dama...spie00}.

\noindent Let us estimate the upper limit of curvature radius $(r_{curv})_{m}
$ (which is defined by capillary/system of capillaries bending), at which
the wave behaviors are displayed under propagation of radiation in channels
\cite{da-rep92}, by considering a photon with the wave vector ${\bf k}$
channeling into capillary with curvature radius $(r_{curv})_{i}$ ($i$-th
layer of capillaries). At small glancing angles, $\theta $, the change of
the longitudinal wave vector, $k_{||}$, under reflection from a capillary
wall is negligibly small; but mainly one changes the transverse wave vector,
$k{\bf _{\perp }}$,

\begin{equation}
k_{\perp}\simeq k\theta\;(\theta<\theta_{c}).
\end{equation}

\noindent Correspondingly, from this relation it follows that the transverse
wavelength will much exceed the longitudinal wavelength that provides the
interference effects to be observable even for very short wavelengths.
Indeed,

\begin{equation}
\lambda_{\perp}=\lambda/\theta>>\lambda
\end{equation}
quantum mechanical principles say that, in order to display the wave
properties of a channeling photon, it is necessary that typical sizes of an
''effective channel'' $\delta_{i}$ , in which waves have been propagating,
be commensurable with the transverse wavelength, i.e. $\delta_{i}(\theta
)\simeq\lambda_{\perp}(\theta)$ (Fig. 2). This condition may be rewritten in
the following form:

\begin{equation}
(r_{curv})_{i}\theta^{3}\sim\lambda\ ,  \label{b-cond}
\end{equation}

\noindent from which we obtain $(r_{curv})_{m}\sim 10\,\ cm$ for a photon of
$\lambda \sim 1\ $\AA\ wavelength. So, from this simple estimate we can
conclude that the relation (\ref{b-cond}) provides a specific dependence for
surface bound state propagation of X-rays - surface channeling - along the
curved surfaces (for instance, in capillary systems) (see also \cite
{liu...prl97}).

\section{PROPAGATION EQUATION}

\subsection{Surface channeling}

\noindent Since the waveguide is a hollow cylindric tube, if the absorption
is considered to be negligible, the interaction potential, in which a wave
propagates, is determined by Eq.(\ref{e-pot-curv}) \noindent with the
radiation polarizability parameter $\delta_{0}\simeq\theta_{c}^{2}$. Solving
the wave equation we are mainly interested in the surface propagation,
which, in fact, defines a wave guiding character inside the channel ($%
r_{\perp}\simeq r_{1},\ \rho\ll r_{1}$) \cite{alda-nim98}

\[
E\left( r\right) \simeq\sum_{m}C_{m}u_{m}(\rho)\ e^{i\left( kz+m\varphi
\right) }\ ,
\]

\begin{equation}
u_{m}(\rho)\propto\left\{
\begin{array}{rcl}
Ai_{m}(\rho) & , & \rho>0 \\
\alpha{Ai^{\prime}}_{m}(0)\,e^{\alpha\rho} & , & \rho<0\,\,\,\,\,\,(\alpha>0)
\end{array}
\right.  \label{a}
\end{equation}

\noindent where $Ai_{m}(x)$ is the Airy function, and $\alpha$ is the
arbitrary unit characterizing the capillary substance. Evidently, these
expressions are valid only for the lower-order modes and in the vicinity of
a channel surface. \noindent The expression (\ref{a}) characterizes the
waves that propagate close to the waveguide wall, or in other words, the
equation describes the grazing modal structure of the electromagnetic field
inside a capillary (surface bound X-ray channeling states) (Fig. 3). The
solution shows also that the wave functions are damped both inside the
channel wall and going from the wall towards the center. It should be
underlined here that the bound modal propagation takes place without the
wave front distortion. The analysis of these expressions allows us also to
conclude that almost all radiation power is concentrated in the hollow
region and, as a consequence, a small attenuation along the waveguide walls
is observed.

\noindent\noindent As for the supported modes of the electromagnetic field,
estimating a characteristic radial size of the main grazing mode ($m=0$)
results in
\begin{equation}
\overline{u}_{0}\ \simeq\sqrt[3]{\frac{\lambda^{2}r_{1}}{2\pi^{2}}}\quad,
\end{equation}

\noindent and we can conclude that the typical radial size $\overline{u}_{0}$
may overcome the wavelength $\lambda$, whereas the curvature radius $r_{1}$
in the trajectory plane exceeds the inner channel radius, $r_{0}$: $%
\overline {u}_{0}\gg\lambda$ (for example, $\overline{u}_{0}\gtrsim0.1\ \mu$%
m for a capillary channel with the radius $r_{0}=10\ \mu$m).

\subsection{\noindent Bulk channeling}

\noindent Above we have considered the transmission of X-ray beams by
capillary systems of micron- and submicron-size channels. Obviously, in that
case we deal with surface channeling of radiation due to the fact that the
channel sizes are larger than the radiation wavelength at least by three
orders of magnitude. However, the situation sharply varies in the case when
the sizes of channels become comparable with the radiation wavelength. In
practice it means, that the angle of diffraction for the given wave,
determined as $\theta _{d}=\lambda /d$ (being $d$ a capillary diameter),
becomes comparable with a critical angle of total external reflection. In
other words, the transverse wavelength of a photon approaches the diameter
of a capillary: $\lambda _{\perp }/d\sim 1$. In this case channeling of
photons (note, not superficial (surface) channeling!) in channels of
capillary systems should occur, i.e. we actually deal with a X-ray waveguide
optics similar to a light waveguide one. Under the condition of the ordering
channels in the system cross section, the capillary nanostructures are
similar to crystals in relation to the charged particles, flying by under
small angles to the main crystallographic directions.

\noindent As the analysis of the wave equation shows, the task cannot be
analytically resolved for the real nanotube potential. For the sake of
simplicity, let us consider the problem in the radial approximation for the
periodic field of a multilayer waveguide with the size $d_{0}$ of a central
channel and the distance $d$ between the layers composing a waveguide wall.
An interaction potential of the radiation in such a waveguide system may be
presented as follows:

\begin{equation}
V\left( r\right) =\sum_{n}V_{n}\left( r\right) =k_{r}^{2}\left[
1+\Delta\sum_{n}\delta\left( \left| r\right| -\frac{d_{0}}{2}-nd\right) %
\right] \ ,  \label{v-period}
\end{equation}

\noindent where $\Delta\equiv\overline{\delta}_{0}d$ is the spatially
averaged polarizability of the wall substance.

\noindent Taking into account the boundary conditions and because of the
potential symmetry, one may conclude that for the central channel $\left|
r\right| \leqslant d_{0}/2$ the solution of the propagation equation in the
transverse plane will be defined by the simple expression

\begin{equation}
E_{0}(r)=\left\{
\begin{array}{rcl}
a\ \cos(k_{r}r) & , & \text{even mode} \\
a\ \sin(k_{r}r) & , & \text{odd mode}
\end{array}
\right.  \label{e-central}
\end{equation}

\noindent Then we define the equation solution for the 1st layer $%
d_{0}/2\leqslant\left| r\right| \leqslant d_{0}/2+d$ by superposition of the
opposite-directed waves $E(r)=b\ e^{ik_{r}r}+c\ e^{-ik_{r}r}$. As it has
been done in the previous section, we make the mathematical assumption that
all the modes of the total energy operator constitute a complete set of
functions in the sense that an arbitrary continuous function can be expanded
in terms of them. Then, we have a wave function $E(r)$ at a particular
instant in time that obeys the continuous boundary conditions at the walls.
In other words, we impose on the solutions the requirements that the wave
function $E$ and its transverse derivative $E_{r}^{^{\prime}}$ be continuous
at the wall-channel boundary

\begin{equation}
\left\{
\begin{array}{rcl}
E\left(
{\displaystyle{d_{0} \over 2}}%
+d-0\right) =E\left(
{\displaystyle{d_{0} \over 2}}%
+d+0\right) & , &  \\
E_{r}^{^{\prime}}\left(
{\displaystyle{d_{0} \over 2}}%
+d\right) =e^{i\varkappa d}\ E_{r}^{^{\prime}}\left(
{\displaystyle{d_{0} \over 2}}%
\right) & , &
\end{array}
\right.  \label{cond-continuum}
\end{equation}

\noindent where $\varkappa$ is the quasimomentum determined by the Bloch
theorem: $E\left( r+d\right) =e^{i\varkappa d}E(r)$ - for the periodical
potential function $V\left( r\right) =V\left( r+d\right) $. From these
expressions we obtain the dispersion relations for even and odd waves
\begin{equation}
\left(
{\displaystyle{\tan%
{\displaystyle{k_{r}d_{0} \over 2}} \atopwithdelims.. \cot%
{\displaystyle{k_{r}d_{0} \over 2}}}}%
\right) =\left(
{\displaystyle{-%
{\displaystyle{k^{2}\Delta \over k_{r}}}+%
{\displaystyle{\cos\left( k_{r}d\right) \ -\ e^{i\varkappa d} \over \sin\left( k_{r}d\right) }} \atopwithdelims.. %
{\displaystyle{k^{2}\Delta \over k_{r}}}-%
{\displaystyle{\cos\left( k_{r}d\right) \ -\ e^{i\varkappa d} \over \sin\left( k_{r}d\right) }}}}%
\right)  \label{dispers}
\end{equation}

\noindent that allow the eigenvalue/eigenfunction problem to be solved.

\noindent Now it is interesting to write the wave functions of the supported
modes for the narrow channel $\left\{ k_{r}d_{0},k_{r}d\right\} \ll1$

\begin{equation}
E_{n}(r)\propto\left\{
\begin{array}{rcl}
\cos(k_{r}r)\ e^{ikz} & , & -%
{\displaystyle{d_{0} \over 2}}%
\leqslant\left| r\right| \leqslant%
{\displaystyle{d_{0} \over 2}}%
\\
\cos%
{\displaystyle{k_{r}d_{0} \over 2}}%
\
{\displaystyle{e^{i\varkappa r}\sin\left( k_{r}\left| \widetilde{r}\right| \right) -\sin\left[ k_{r}\left( \left| \widetilde {r}\right| -d\right) \right]  \over \sin\left( k_{r}d\right) }}%
\ e^{i\left( \varkappa nd+kz\right) } & , & -%
{\displaystyle{d_{0} \over 2}}%
+nd\leqslant\left| r\right| \leqslant%
{\displaystyle{d_{0} \over 2}}%
+\left( n+1\right) d
\end{array}
\right. ,  \label{e-narrow}
\end{equation}

\noindent where $\left| \widetilde{r}\right| \equiv \left| r\right|
-d_{0}/2-nd$. In this case we see that the Eqs.(\ref{dispers}) may be solved
only for even modes. However, it is more important to underline that the
even mode exists for any ratio between the channel size and the layer
distance. The spatial distribution of the mode has a maximum at the centre
of the channel, and due to the leak through the potential barrier of wall
layers we observe the propagation of radiation in substance. The radiation
intensity for the successive layer decreases following an exponential law
and is characterized by a local maximum far from the layer wall (the radial
wave function distribution for the case of quasidistant layer system is
shown in Fig. 4).

\bigskip

\noindent Because of the small wall thicknesses of nanotube channels (less
than $\lambda_{\perp}\lesssim100$ \AA) we have to note that part of the
radiation, channeling inside a nanotube structure, will undergo
''tunneling'' through the potential wall barrier. A simple analysis of the
radiation propagation in systems both for the case of macroscopic channel
and for the case of totally isotropic spatial structure, shows the presence
of the main channeling mode (the main bound state) for any structure,
whereas the high modes may be suppressed for specific channel sizes. Hence,
nanotubes present a special interest as waveguides, which allow the
supported modes to be governed. Moreover, there is a special interest in
studying the dispersion of radiation in a nanosystem with a multilayered
wall. As follows from the analysis of the general equation of radiation
propagation considered above, at any correlation between the channel size
and the interlayer distance at least one mode (bound state) should be formed
in such a structure. In that case the diffraction of waves reflected from
various layers of the channel wall should be observed, hence affecting the
radiation distribution at the exit of system.

\noindent Evidently, the efficiency of these structures for applications
have to be analyzed, despite the importance of the nanotube X-ray waveguide
phenomenon from the fundamental point of view. \noindent The problems
associated with X-ray and neutron channeling in capillary nanotubes (single-
and multi-wall systems) present a special interest and will be analyzed in a
subsequent paper.

\noindent

\section{\bf CONCLUSION}

\noindent The reduction in\ the channel size of capillary structures, as
well as the discovery of a new class of natural nanosystems - i.e. carbon
nanotubes - puts the problem of passage of X-ray quanta through these
systems on a new qualitative level.

\noindent In the present work the general unified theory, allowing us to
describe processes of radiation passage through capillary systems in a wide
range of channel diameters, is offered. Our analysis shows, that in the case
of micron diameters the surface channeling of quanta presents in the
mechanism of propagation (the part of radiation is distributed, being
trapped by the bent surface of the channel), that can influence essentially
the angular radiation distribution behind capillary systems. A decrease in
size up to the nano-level results in a transition from surface channeling to
bulk channeling. Thus, all radiation is involved in the process of the modal
propagation, as opposed to the surface channeling when only part of the
radiation is subject to the bound spreading.

\noindent Recently, it has been demonstrated that for specific cases a
decrease in the angular divergence of X radiation after passing through
capillary systems may be observed \cite{ca-da...apl01}. We have found strong
differences between the observed and expected FWHM (full width at half
maximum) values. From the general ray approach estimations, a FWHM value of
at least $2\theta_{c}\simeq0.9%
{{}^\circ}%
$ for $4\ $keV synchrotron radiation photons is expected. This width exceeds
both the experimental (FWHM$_{\exp}^{SR}$[4\ keV] $=0.28%
{{}^\circ}%
$) and Gaussian fitted (FWHM$_{G}^{SR}$[4\ keV] $=0.22%
{{}^\circ}%
$) values.\ An analogous feature takes place also in the case of harder
radiation. Such a behavior indicates an essential redistribution of radiation
scattered inside the channels. The obtained discrepancy in experimental and
theoretical results is not reproduced within the framework of\ the ray
approximation and may be explained using the wave approach method, in order
to describe the angular distribution of the reflected beam.

\noindent The discovery of a new class of ordered systems like nanostructures
(fullerenes, nanotubes) opens up interesting opportunities for studying
coherent effects of the radiation interacting with nanosystems. The interest is
not limited to fundamental research, on the contrary, there is a big potential
for using the nanosystem samples, in order to develop new technological ideas
(nanotube benders, nanosystem detectors, nanosystems as a source of
electromagnetic radiation, nanotube undulators, etc.). For instance, recently
new types of undulators for new generations of X-ray sources (FEL) have been
studied, based on the manufacturing of a crystal undulator obtained
by making a periodic series of micro trenches on the crystal surface (stripe
distortion potential) \cite{BBDGetal}, a research that which might result in the
future creation of undulators based on nanostructures.

\noindent The main purpose of the present work is fundamental - i.e. to study the
processes accompanying the X-ray transmission by capillary structures - in
spite of the fact that capillary optics applications are covering larger and
larger areas in X-ray physics and chemistry, biology and medicine. X-ray and
neutron research activities over the last ten years demonstrated that
capillary optics is a powerful instrument to guide neutral particle beams.
Capillary/polycapillary optics can be applied in numerous branches of X-ray
research, e.g. spectroscopy, fluorescence analysis, \ crystallography,
imaging technics, tomography, lithography, etc. \cite{kum-selpap00}.

\section*{ACKNOWLEDGMENTS}

\noindent This work was partly supported by the NANO experiment of the
Commissione Nazionale V of the Istituto Nazionale di Fisica Nucleare and by the
Russian Federation Federal Program ''Integration''.

\newpage

Figure captions

Figure 1: The change of the interaction potential between the flat surface
(A.) and the curved one (B.). For simplicity, in calculations the real
potential (B.) may be replaced by the model potential (C.).

Figure 2: Illustration of X-ray reflection from the inner capillary surface.
At glancing angles $\theta,$ when the cross size of a beam $%
\delta_{i}(\theta)$\ becomes comparable with the transverse wavelength $%
\lambda_{\perp }(\theta)$, the radiation is grasped in a mode of surface
channeling.

Figure 3: The radial distributions of the main bound mode of radiation
inside a capillary channel for various channel diameters. The decrease of
diameter ($2r_{0}$) results in a spatial displacement of the distribution away
from the channel wall towards the center. The wall surface position is shown
by the dotted line.

Figure 4: The typical radial wave function distribution for a periodic
multilayer waveguide, where $d$ is the distance between the layers composing
a waveguide wall.

\bigskip

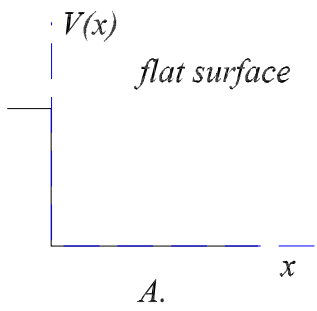
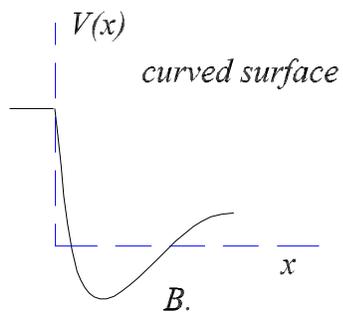
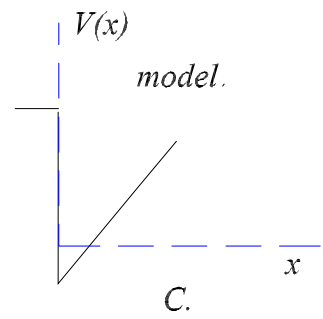
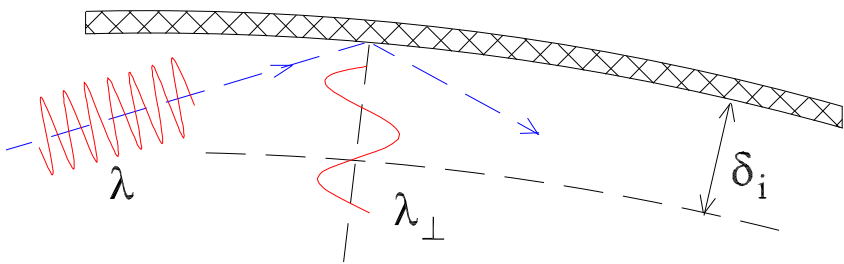
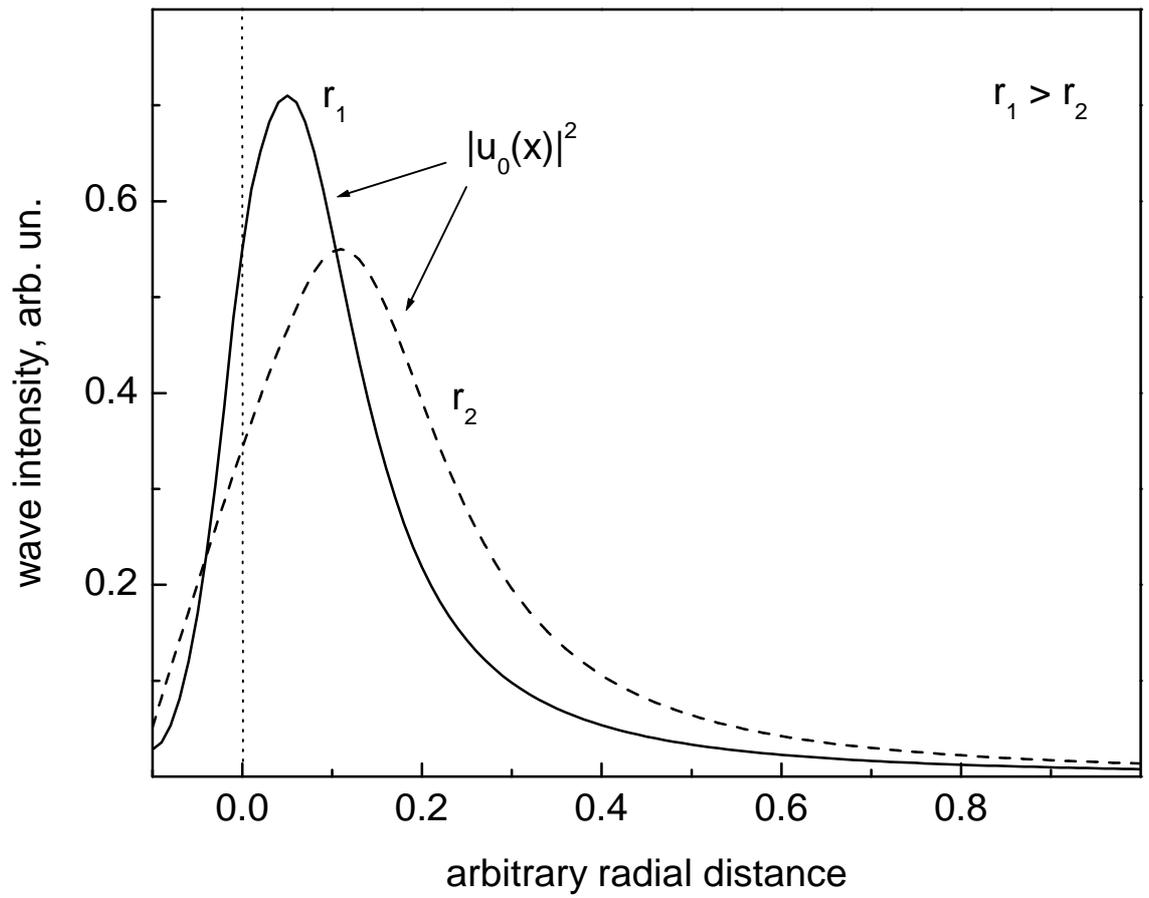
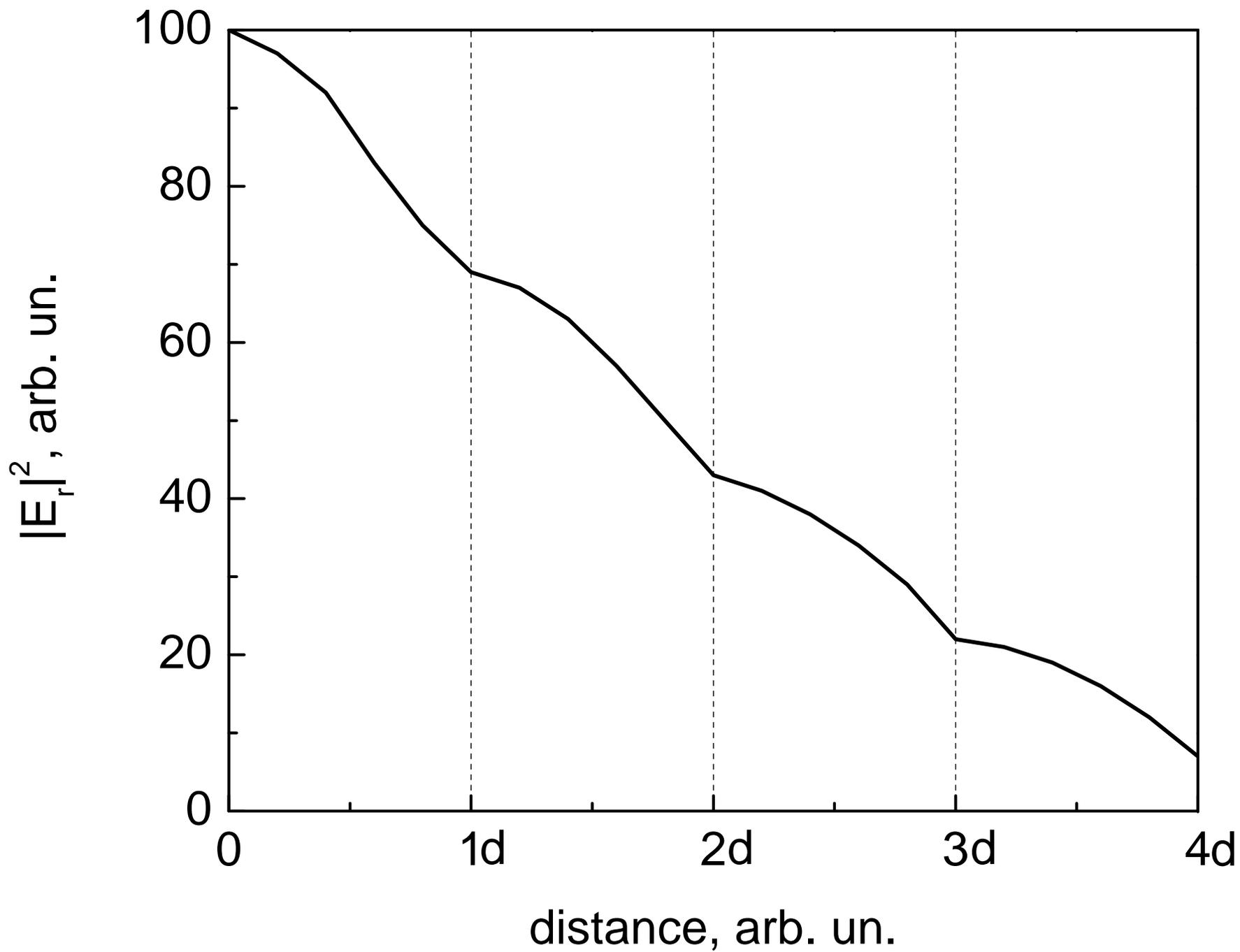
\end{document}